\documentclass[prc,twoside,superscriptaddress,showkeys,showpacs, twocolumn]{revtex4-1}%

\usepackage{graphicx, latexsym, amssymb, amsmath, color, multirow, mathrsfs}
\usepackage[section]{placeins}
\usepackage{amsmath}
\usepackage{amsfonts}
\usepackage{amssymb}
\usepackage{graphicx}
\newcommand{\beq}{\begin{equation}}
\newcommand{\eeq}{\end{equation}}
\newcommand{\beqn}{\begin{eqnarray}}
\newcommand{\eeqn}{\end{eqnarray}}
\newcommand{\bea}{\begin{array}}
\newcommand{\eea}{\end{array}}
\newcommand{\bsub}{\begin{subequations}}
\newcommand{\esub}{\end{subequations}}
\newcommand{\bpm}{\begin{pmatrix}}
\newcommand{\epm}{\end{pmatrix}}

\newcommand{\scr}[1]{{\mathscr #1}}
\newcommand{\ff}[1]{\frac{1}{#1}}

\newcommand{\lrl}[1]{\left|#1\right|}

\newcommand{\lrb}[1]{\left(#1\right)}
\newcommand{\lrs}[1]{\left[#1\right]}

\newcommand{\svec}[1]{{\mbox{\boldmath${ #1}$}}}
\newcommand{\ivec}{\vec}
\newcommand{\re}{\nonumber\\}

\begin{document}

\title{Relativistic Hartree-Fock-Bogoliubov theory with Density Dependent Meson-Nucleon Couplings}

\author{Wen Hui Long}\email{whlong@pku.org.cn}
\affiliation{School of Nuclear Science and Technology, Lanzhou University, 730000 Lanzhou, China}
\affiliation{Physik-Department der Technischen Universit\"at M\"unchen, D-85748 Garching, Germany}
\affiliation{School of Physics and State Key Laboratory of Nuclear Physics and Technology, Peking University, 100871 Beijing, China}
\author{Peter Ring}
\affiliation{Physik-Department der Technischen Universit\"at M\"unchen, D-85748 Garching, Germany}
\author{Nguyen Van Giai}
\affiliation{CNRS-IN2P3, UMR 8608, F-91406 Orsay Cedex, France}
\affiliation{Univ Paris-Sud, F-91405 Orsay, France}
\author{Jie Meng}
\affiliation{School of Physics and State Key Laboratory of Nuclear Physics and Technology, Peking University, 100871 Beijing, China}
\date{\today}

\begin{abstract}
Relativistic Hartree-Fock-Bogoliubov (RHFB) theory with density-dependent meson-nucleon couplings is presented. The
integro-differential RHFB equations are solved by expanding the different components of the quasi-particle spinors in the complete set of eigen-solutions of the Dirac equations with Woods-Saxon potentials. Using the finite-range Gogny force D1S as an effective interaction in the pairing channel, systematic RHFB calculations are performed for Sn isotopes and $N=82$ isotones. It is demonstrated that an appropriate description of both mean field and pairing effects can be obtained within RHFB theory with finite range Gogny pairing forces. Better systematics are also found in the regions from the stable to the neutron-rich side with the inclusion of Fock terms, especially in the presence of $\rho$-tensor couplings.
\end{abstract}

\pacs{24.10.Cn,  24.10.Jv   21.60.Jz,  21.30.Fe, }

\keywords{RHFB theory; DDRHF theory; RHF method; Pairing correlations; Dirac Woods-Saxon basis}
\maketitle

\section{Introduction}
During the past decades, much success has been achieved in nuclear physics by relativistic density functional theories. One of the most outstanding candidates is the relativistic Hartree approach with the no-sea approximation, namely the relativistic mean field (RMF) theory~\cite{Walecka:1974, Serot:1986, Reinhard:1989, Ring:1996, Bender:2003, Vretenar:2005, Meng:2006}. Within the RMF framework, valuable information has been obtained for the structure of the
nuclei in and far from the valley of $\beta $-stability, including both for ground states~\cite{Reinhard:1989, Ring:1996, Meng:2006} and excited states~\cite{Vretenar:2005, Paar:2007}. At the same time, considerable effort has been devoted to relativistic Hartree-Fock (RHF) theory~\cite{Brockmann:1978, Horowitz:1984, Bielajew:1984, Blunden:1987, Bouyssy:1985, Bouyssy:1987, Bernardos:1993, Marcos:2004}. However, because of its numerical complexity, for a long time, it failed in a quantitative description of nuclear systems. Only in the recent years, with the growth of computational facilities and the development of new methods, density dependent relativistic Hartree-Fock (DDRHF) theory has shown significant improvements in a quantitative description of nuclear phenomena~\cite{Long:2006, Long:2006PS, Long:2007, Long:2008, Liang:2008, BYSun:2008} with a similar accuracy as RMF.

In DDRHF, the Lorentz covariant structure is kept in full rigor, which guarantees the self-consistent determination of the spin-orbit coupling~\cite{Long:2006} and all well-conserved relativistic symmetries, e.g., the pseudo-spin symmetry in the nuclear spectrum~\cite{Long:2006PS}. In addition, significant improvements on the relativistic description of shell structures have been gained with the newly introduced constituents by the Fock terms, i.e., the pseudo-vector pion and the $\rho$-tensor couplings. In {Refs.~\cite{Long:2008,Lalazissis:2009}}, the consistency of the evolution of the shell structure has been considerably improved by the pion exchange potential, in fact, by its tensor part. With the inclusion of $\rho$-tensor couplings, the common disease of several artificial shell closures existing in the RMF calculations~\cite{Geng:2006}, has been cured in DDRHF theory and the pseudo-spin symmetry is also better preserved~\cite{Long:2007}. {Besides the Fock terms in the isovector channel, those derived from} the isoscalar $\sigma$ and $\omega$ couplings are found to play {a dominant role in reproducing the} characteristic experimental
$Z$-dependence of the spin-orbit splitting around the sub-shell closure $Z=64$~\cite{Long:2009}. It has also been demonstrated that the isoscalar Fock terms are essential for self-consistent description of the spin-isospin resonances within RPA~\cite{Liang:2008} and prediction of neutron star properties~\cite{BYSun:2008}.

On the other hand, the development of the radioactive ion beam (RIB) facilities~\cite{Bertulani:2001} has opened a new frontier for nuclear physics, the field of exotic nuclei far from the valley of stability~\cite{Mueller:1993, Tan:1995, Hansen:1995, Casten:2000, Mul:2001, Jensen:2004, Jonson:2004} {and the} upgrades and constructions of the RIB facilities \cite{Zhan04, Henning04, Yano04, riaweb} in recent few years will provide us with new possibilities to
study exotic modes in nuclear systems. The current application of DDRHF is limited {to nuclei} in the $\beta $-stability valley and pairing effects {in open} shell nuclei are treated only within the BCS approximation \cite{Long:2006, Long:2007}. In weakly bound systems like exotic nuclei close to the drip lines, the Fermi surface
of one type of nucleons is close to the particle continuum, and the single nucleon separation energies are comparable to the pairing gaps. This results in an enhancement of scattering of Cooper pairs into the continuum {due to pairing} correlations. Thus, it becomes necessary to include the continuous part of the single-particle spectrum to describe the unstable nuclei.

It is now the general consensus that a unified and self-consistent description of both mean field and pairing correlations can be obtained with the Bogoliubov transformation and automatically the continuum effects are efficiently taken into account~\cite{Meng:1996, Meng:1998a}. {In this manuscript relativistic Hartree-Fock-Bogoliubov (RHFB) theory with density dependent meson-nucleon couplings is presented as a natural extension of DDRHF. Its} content is organized as follows. In Section \ref{sec:RHFB} {we introduce} the general formalism of the RHFB theory with both
zero-range (delta) and finite range (Gogny) pairing forces, where the integro-differential RHFB equations are solved by expanding the lower and upper components of the quasi-particle spinors on the complete set of solutions of the Dirac equation with a Woods-Saxon (DWS) type potential. {The comparison between different treatments of pairing correlations is discussed in Section \ref{sec:Results} and systematic RHFB calculations are performed and discussed by taking Sn isotopes and $N=82$ isotones as representative cases}. Finally, a brief summary is given in Section \ref{sec:Conclusions}.

\section{General Formalism}\label{sec:RHFB}
We briefly recall here the general features of DDRHF theory in order to make understandable its generalization to the RHFB case. More details can be found in Refs.~\cite{Bouyssy:1987, Long:2006, Long:2007} whereas the effective interactions used in this work have been introduced in Refs.~\cite{Long:2006, Long:2007, Long:2008}.

\subsection{Energy Functional and Relativistic Hartree-Fock potentials}
As generally recognized, the nucleon-nucleon interaction is mediated by the exchange of mesons {with isoscalar and isovector character}. The understanding of nuclear structure at the microscopic level, therefore, has to be achieved in the same language. Consistent with this criterion, the model Lagrangian, i.e., the theoretical starting point, contains the degrees of freedom associated with the nucleon, the $\sigma$-, $\omega$-, $\rho$-, and $\pi$-meson fields, and the photon field ($A$)~\cite{Bouyssy:1987}. Following the standard variational procedure of the  Lagrangian \cite{Serot:1986, Bouyssy:1987}, one {finds} the equations of motion for mesons, nucleons and photons, namely the Klein-Gordon, Dirac and Proca equations, and the continuity equation, i.e., the energy-momentum conservation relation, from which is derived the Hamiltonian of the system. In terms of the creation and annihilation operators ($c_\alpha^\dag, c_\alpha $) defined by the stationary solutions of the Dirac equation, the Hamiltonian can be generally expressed as,
 \begin{equation}\label{Hamiltonian}
H = \sum_{\alpha\beta} c_\alpha^\dag c^{}_\beta T_{\alpha\beta} +
\ff2 \sum_{\alpha\alpha'\beta\beta'} c_\alpha^\dag c_\beta^\dag
c^{}_{\beta'}c^{}_{\alpha'} \sum_\phi
V_{\alpha\beta\beta'\alpha'}^\phi,
 \end{equation}
where $T_{\alpha\beta}$ represents the kinetic energy and the two-body terms $ V_{\alpha\beta\beta'\alpha'}^\phi$ correspond to different types of meson (or photon) nucleon couplings denoted by $\phi$,
 \begin{align}
T_{\alpha\beta} = &\int d\svec x \bar\psi_{\alpha}(\svec x) \lrb{-i\svec\gamma\centerdot\svec
\nabla +M}\psi_\beta \\
V_{\alpha\beta\beta'\alpha'}^\phi = & \int d\svec x d\svec x' \bar
\psi_\alpha(\svec x) \bar \psi_\beta(\svec x')
\Gamma_{\phi}(x,x')\re&~~~~~~~~~~~~~~~\times D_{\phi}(\svec x,\svec
x')\psi_{\beta'}(\svec x')\psi_{\alpha'}(\svec x).
 \end{align}

In the two-body interaction terms, the {interaction matrices} $\Gamma_{\phi}(x,x^{\prime})$ read as
 \bsub\label{INDEX}\begin{align}
\Gamma_{\sigma}  (\svec x,\svec x') \equiv &  -g_{\sigma}(\svec x)g_{\sigma}(\svec x'),\\ %
\Gamma_{\omega}  (\svec x,\svec x') \equiv &  \lrb{g_{\omega}\gamma_{\mu}}_{\svec x}\lrb{g_{\omega}\gamma^{\mu}}_{\svec x'},\\ %
\Gamma_{\rho}^{V}(\svec x,\svec x') \equiv &  \lrb{g_{\rho}\gamma_{\mu}\vec {\tau}}_{\svec x}\cdot %
                                              \lrb{g_{\rho}\gamma^{\mu}\vec{\tau}}_{\svec x'},\\ %
\Gamma_{\rho}^{T}(\svec x,\svec x') \equiv &  \frac{1}{4M^{2}}\lrb{f_{\rho}\sigma_{\nu k}\vec{\tau}\partial^{k}}_{\svec x}\cdot%
                                              \lrb{  f_{\rho}\sigma^{\nu l}\vec{\tau}\partial_{l}}_{\svec x'},\\%
\Gamma_\rho^{VT}(\svec x,\svec x')  \equiv & \ff{2M}\lrb{f_\rho\sigma^{k\nu}\ivec\tau\partial_k}_{\svec x}\cdot%
                                                    \lrb{g_\rho\gamma_\nu\ivec\tau}_{\svec x'}\re & + %
                                             \ff{2M}\lrb{g_\rho\gamma_\nu\ivec\tau}_{\svec x}\cdot %
                                                    \lrb{f_\rho\sigma^{k\nu}\ivec\tau\partial_k}_{\svec x'},\\%
\Gamma_{\pi}(\svec x,\svec x')      \equiv & \frac{-1}{m_{\pi}^{2}}\lrb{f_\pi\ivec\tau\gamma_5\gamma_\mu\partial^\mu}_{\svec x}%
                                                              \cdot\lrb{f_\pi\ivec\tau\gamma_5\gamma_\nu\partial^\nu}_{\svec x'},\\%
\Gamma_{A}(\svec x,\svec x')        \equiv & \frac{e^{2}}{4}\lrb{\gamma_\mu(1-\tau_3)}_{\svec x}%
                                                            \lrb{\gamma^\mu(1-\tau_3)}_{\svec x'}.
\end{align} \esub
In coordinate space, the propagators $D_{\phi}(\svec x,\svec x^{\prime})$ for the meson fields have a Yukawa form
 \begin{equation}
D_{\phi}(\svec x,\svec x')=\frac {1}{4\pi}\frac{e^{-m_{\phi}\lrl{\svec x-\svec x'}}}{\lrl{\svec
x-\svec x'}}. \label{Yukawa}
 \end{equation}
For the photon field, the propagator $D_{A}(\svec x, \svec x')$ can be written as,
 \begin{equation}
D_{A}(\svec x, \svec x')=\frac {1}{4\pi}\frac{1}{\lrl{\svec x-\svec x'}}.
 \end{equation}
In the above expressions (Eqs.(2-5)), $M$ denotes the {nucleon mass} and $m_{\sigma}$ ($g_{\sigma}$), $m_{\omega}$ ($g_{\omega}$), $m_{\rho}$ ($g_{\rho},f_{\rho}$), and $m_{\pi}$ ($f_{\pi}$) are the masses (coupling constants) corresponding to $\sigma$, $\omega$, $\rho$, and $\pi$ mesons. In this paper, we use arrows to denote isospin vectors and bold types for vectors in coordinate space.

In the Hamiltonian (\ref{Hamiltonian}),  the indices $\alpha$, $\beta$, $\alpha'$, $\beta'$ run over all the {single-particle states} ($\psi_\alpha$) with positive energies ($\alpha=k$) and negative energies ($\alpha=l$). As it is commonly done in the mean field approach, the so-called \emph{no sea} approximation is adopted and the contributions from the negative energy states are neglected. Then, the energy functional can be obtained from the following expectation value
\begin{equation}
E=\left\langle \Phi_{0}\right\vert H\left\vert \Phi_{0}\right\rangle , \label{Energy-Functional}
\end{equation}
where $\left\vert \Phi_0\right\rangle$ is the Hartree-Fock ground state in the no-sea approximation~\cite{Bouyssy:1987}. In the energy functional (\ref{Energy-Functional}), the contributions of the two-body interactions $V_{\phi}$ consist of two parts, the direct (Hartree) and exchange (Fock) terms. With only the direct contributions, Eq. (\ref{Energy-Functional}) leads to the energy functional of the RMF theory. With both direct and exchange
contributions we obtain the energy functional for the DDRHF theory.

In {spherically symmetric systems} the Dirac spinor can be written as,
\begin{equation}
\psi_{\alpha}({\mbox{\boldmath${ r}$}})=\frac{1}{r}\left(
\begin{array}
[c]{c}
iG_{a}(r){\mathcal{Y}}_{j_{a}m_{a}}^{l_{a}}(\hat{{\mbox{\boldmath${ r}$}}})\\[0.5em]
-F_{a}(r){\mathcal{Y}}_{j_{a}m_{a}}^{l_{a}^{\prime}}(\hat
{{\mbox{\boldmath${ r}$}}})
\end{array}
\right)  . \label{Spinor}
\end{equation}
The radial wave functions $G_{a}(r)$ and $F_{a}(r)$ characterize the upper (large) and lower (small) components and
${\mathcal{Y}}_{jm}^{l}$ are the {spherical harmonic spinors}. Here, the sub-index $\alpha=\left\{  a,m_{a}\right\} =\left\{ n_{a},l_{a},l_{a}^{\prime},j_{a},m_{a}\right\} $ contains the quantum numbers $n_{a}$ (number of nodes of the upper component $G_{a}$), $j_{a},m_{a}$ (total angular momentum and its projection to the $z$-{axis}), and $l_{a},l_{a}^{\prime}$ \ (orbital angular momenta with $l_{a}+l_{a}^{\prime } =2j_{a}$). In the following, we will use
the Latin indices for the {sub-set} $\left\{ nll^{\prime}j\right\}  $ and Greek indices for the full set $\left\{ njll^{\prime}m\right\}$.

By taking the variation of the energy functional (\ref{Energy-Functional}) with respect to the Dirac spinor
(\ref{Spinor}), we obtain the {spherical} Dirac Hartree-Fock equation as
 \beq\label{Spherical}
\int d\svec r' h(\svec r, \svec r') \psi(\svec r')  = \varepsilon\psi(\svec r),
 \eeq
where $\varepsilon$ is the single-particle energy ({including the rest mass}) and the single-particle Dirac Hamiltonian $ h(\svec r, \svec r')$ contains the kinetic energy $h^{\text{kin}}$, the direct {local potential $h^{\text{D}}$ and exchange non-local potential $h^{\text{E}}$,}
  \bsub\label{HamiltonianS}\begin{align}
&h^{\text{kin}}(\svec r, \svec r') = \lrs{\svec\alpha\cdot\svec p +
\beta M }\delta(\svec r - \svec r'),\\
&h^{\text{D}}(\svec r, \svec r') = \lrs{\Sigma_T(\svec r)\gamma_5 + \Sigma_0(\svec
r) + \beta\Sigma_S(\svec r)}\delta(\svec r - \svec r'),\\
&h^{\text{E}}(\svec r, \svec r') =  \lrb{\bea{cc}Y_G(\svec r, \svec r') &Y_F(\svec r, \svec r')\\[0.5em]
X_G(\svec r, \svec r')&X_F(\svec r, \svec r')\eea}.
 \end{align}\esub
In the above expression, the {local} self-energies $\Sigma_{S}$, $\Sigma_{0}$ and $\Sigma_{T}$ contain the contributions from the direct {(Hartree)} terms~\cite{Serot:1986, Reinhard:1989, Ring:1996, Meng:2006} and the rearrangement terms~\cite{Vretenar:2005}. {The non-local self-energies $X_G$, $X_F$, $Y_G$ and $Y_F$ come from the exchange (Fock) terms and they take the general form}
 \bsub\label{Non-local}\begin{align}
X_{G_a}^{(\phi)}(r,r') =&\sum_b\scr T_{ab}^{\phi}\hat j_b^2\lrb{g_\phi F_b}_r \scr
R_{ab}^{X_G}(m_{\phi};r,r^{\prime})\lrb{ g_\phi G_b}_{r'},\\
X_{F_a}^{(\phi)}(r,r') =&\sum_b\scr T_{ab}^{\phi}\hat j_b^2\lrb{g_\phi F_b}_r \scr
R_{ab}^{X_F}(m_{\phi};r,r^{\prime})\lrb{ g_\phi F_b}_{r'},\\
Y_{G_a}^{(\phi)}(r,r') =&\sum_b\scr T_{ab}^{\phi}\hat j_b^2\lrb{g_\phi G_b}_r \scr
R_{ab}^{Y_G}(m_{\phi};r,r^{\prime})\lrb{ g_\phi G_b}_{r'},\\
Y_{F_a}^{(\phi)}(r,r') =&\sum_b\scr T_{ab}^{\phi}\hat
j_b^2\lrb{g_\phi G_b}_r \scr
R_{ab}^{Y_F}(m_{\phi};r,r^{\prime})\lrb{ g_\phi F_b}_{r'}.
 \end{align}\esub
In these expressions, $g_\phi$ represents the coupling constants, $\hat {\jmath}_{b}=\sqrt{2j_{b}+1}$, and $\scr T_{ab}^{\phi}$ denotes the isospin factors: $\delta_{\tau_a\tau_b}$ and $2-\delta_{\tau_a\tau_b}$ respectively for isoscalar and isovector channels. For example, one has $\scr R^{Y_G} = \scr R^{X_F} = -\scr R^{Y_F} = -\scr R^{X_G} = \scr R^{(\sigma)}$ for the $\sigma$-scalar coupling, and
 \begin{equation}\label{coprod_sigma}
{\mathscr R}_{ab}(m_{\sigma},r,r^{\prime})=\sum
_{L}^{\prime}\lrb{C_{j_{a}\frac{1}{2}j_{b}-\frac{1}{2}}^{L0}}^2 R_{LL}(m_{\sigma};r,r^{\prime}).
 \end{equation}
The prime on the sum in Eq. (\ref{coprod_sigma}) {indicates that $L+l_{a}+l_{b}$ must be even, and $R_{L_1L_2}$ stands for}
 \begin{align}
R_{L_1L_2}(m_i; r,r')&=\sqrt{\ff{rr'}}\lrs{ I_{L_1+\ff2}(z) K_{L_2+\ff2}(z')\theta(z'-z)\right.\re
& +\left. K_{L_1+\ff2}(z) I_{L_2+\ff2}(z')\theta(z-z')},
 \end{align}
where $z=m_{\phi}r$, $I_{L+\ff2}$ and $K_{L+\ff2}$ are related to the spherical Bessel and Hankel functions. The detailed expressions of all self-energies entering the HF potentials can be found in Ref.~\cite{Bouyssy:1987} excepted for the rearrangement potentials because the couplings there were assumed density independent. Here, the rearrangement potentials are of course included in the calculations. We observe that, in a non-relativistic reduction, the pion pseudo-vector coupling and the $\rho$-tensor coupling lead to central and tensor nucleon-nucleon interactions and therefore, they play a substantial role in determining the spin-orbit splittings and shell evolutions~\cite{Long:2007, Long:2008}.

In realistic applications, one has to consider the nuclear medium effects. Within the RHF approach, some efforts have been devoted to considering the in-medium effects by introducing non-linear self couplings of the $\sigma$ and $\omega$ fields~\cite{Bernardos:1993} or cubic and quadratic terms of the scalar field ($\bar\psi\psi$)~\cite{Marcos:2004}. Instead of the non-linear self couplings, here we assume a density dependence of the meson-nucleon couplings~\cite{Brockmann:1992, Lenske:1995, Fuchs:1995} as we did before \cite{Long:2006, Long:2007}, which looks more coincident with the model Lagrangian.

As shown in Ref.~\cite{Fuchs:1995}, the density dependence in meson-nucleon couplings leads to rearrangement terms
$\Sigma_{R}^{\mu}$ in the self-energy $\Sigma^\mu$ in order to preserve the energy-momentum conservation,
\begin{equation}
\Sigma^\mu\rightarrow\Sigma^\mu+\gamma_{\mu}\Sigma_{R}^{\mu}~.
\end{equation}
For example, the rearrangement term due to the density dependence in $\sigma$-scalar coupling can be written as,
 \beq
\Sigma_R^{(\sigma)}  = \frac{\partial g_\sigma}{\partial \rho_b} \lrs{\rho_s\sigma + \sum_b
\frac{\hat j_b^2}{g_\sigma r^2}\lrb{G_b Y_b^{(\sigma)} + F_bX_b^{(\sigma)}}},
 \eeq
where {$\rho_{s}$ and $\rho_{b}$ are respectively the local scalar and baryonic densities, and the Fock components $X_b^{(\phi)}$ and $Y_b^{(\phi)}$ can be written as,
 \beq\label{XY}
\lrb{\bea{c}Y_b^{(\phi)}\\[0.5em] X_b^{(\phi)}\eea}_r = \int dr'\lrb{\bea{cc}Y_{G_b}^{(\phi)}
&Y_{F_b}^{(\phi)}\\[0.5em] X_{G_b}^{(\phi)}&
X_{F_b}^{(\phi)}\eea}_{(r,r')}\lrb{\bea{c}G_b\\[0.5em] F_b\eea}_{r'}.
 \eeq }

\subsection{Density-dependent Relativistic Hartree-Fock-Bogoliubov theory}
In open shell nuclei, the effects of pairing correlations, which lead to valence particles spreading over the orbits around the Fermi level, have to be taken into account, either in the BCS approximation~\cite{Bardeen:1957} or by the full Bogoliubov theory~\cite{Gorkov58}. In terms of quasi-particles, the Bogoliubov theory unifies the treatment of $ph$- and $pp$-correlations in a self-consistent description of nuclear orbitals~\cite{Ring:1980}. It is specially significant for the exploration in the regions far from the stability where the simple BCS method may break down.  In the relativistic case~\cite{Kucharek:1991, Gonzalez96} earlier investigations within relativistic Hartree Bogoliubov (RHB) theory have shown that the scattering of the Cooper pairs into the continuum plays an important role for the {formation} of the neutron halos~\cite{Meng:1996, Meng:1998PRL}. Within the Bogoliubov scheme, the single-particle basis $\left\{ c_{\alpha},c_{\alpha }^{\dag}\right\} $ and the quasi-particle basis $\left\{ \beta_{\alpha },\beta_{\alpha}^{\dag}\right\}  $ ($\alpha=1,\cdots,M$) are related by the following transformation
\begin{equation}
\left( \begin{array} [c]{c}
c_{\alpha}\\[0.5em] c_{\alpha}^{\dag}
\end{array}
\right)  = {\mathcal{W}}\left(
\begin{array}[c]{c}\beta_{\alpha}\\[0.5em]
\beta_{\alpha}^{\dag}\end{array} \right)  =\left(
\begin{array}[c]{cc}\psi_{U} & \psi_{V}^{\ast}\\[0.5em]
\psi_{V} & \psi_{U}^{\ast}\end{array} \right)  \left(
\begin{array} [c]{c}
\beta_{\alpha}\\[0.5em]\beta_{\alpha}^{\dag}
\end{array}
\right),
\end{equation}
where $\psi_{U}$ and $\psi_{V}$ are the quasi-particle spinors, of the form {of} Eq. (\ref{Spinor}) in the spherical case. The transformation satisfies unitarity
\begin{equation}
{\mathcal{W}}^{\dag}{\mathcal{W}}=1.
\end{equation}
Following the standard procedure of the Bogoliubov transformation~\cite{Gorkov58}, a relativistic Hartree-Fock-Bogoliubov equation can be derived as~\cite{Kucharek:1991},
 \beq\begin{split}\label{RHFBEQ}
\int d\svec r'&\lrb{\bea{cc}h(\svec r,\svec r')-\lambda&\Delta(\svec r,\svec r') \\[0.5em]
\Delta(\svec r,\svec r') & -h(\svec r,\svec r') +\lambda\eea}\\
&~~~~~~~~~~~~\times \lrb{\bea{c}\psi_U(\svec r')\\[0.5em] \psi_V(\svec r')\eea}
= E\lrb{\bea{c}\psi_U(\svec r)\\[0.5em] \psi_V(\svec r)\eea},
 \end{split}\eeq
where the chemical potential $\lambda$ $\ $is introduced to preserve the particle number on the average. {In the single-particle Hamiltonian $h(\svec r,\svec r')$, the retardation effects are neglected as is usually done in mean field calculations.} The pairing potential can be written as,
\begin{equation}
\Delta_{\alpha}({\mbox{\boldmath${ r}$}},{\mbox{\boldmath${ r}$}}^{\prime
})=-\frac{1}{2}\sum_{\beta}V_{\alpha\beta}^{pp}\left( {\mbox{\boldmath${ r}$}},{\mbox{\boldmath${
r}$}}^{\prime}\right) \kappa_{\beta}({\mbox{\boldmath${ r}$}},{\mbox{\boldmath${ r}$}}^{\prime}),
\label{pairingp}
\end{equation}
where the pairing tensor $\kappa$ is
\begin{equation}
\kappa_{\alpha}({\mbox{\boldmath${ r}$}},{\mbox{\boldmath${ r}$}}^{\prime
})=\psi_{V_{\alpha}}({\mbox{\boldmath${ r}$}})^{\ast}\psi_{U_{\alpha}%
}({\mbox{\boldmath${ r}$}}^{\prime}).
\end{equation}
For the pairing interaction $V^{pp}$ in Eq. (\ref{pairingp}), a phenomenological form is adopted as it has been done with great success in RHB theory~\cite{Gonzalez96, Vretenar:2005} and in conventional HFB theory~\cite{Decharge:1980, Dobaczewski:1984}. The pairing force is either taken as a density-dependent two-body force in a zero range limit,
\begin{equation}
V({\mbox{\boldmath${ r}$}},{\mbox{\boldmath${ r}$}}^{\prime})=V_{0} \delta({\mbox{\boldmath${
r}$}}-{\mbox{\boldmath${ r}$}}^{\prime})\frac{1}{4}\left(  1-{\mbox{\boldmath${ \sigma
}$}}\cdot{\mbox{\boldmath${ \sigma}$}}^{\prime}\right) \left(  1-\frac {\rho(r)}{\rho_{0}}\right) ,
\label{Delta}
\end{equation}
with {an adjusted} strength $V_{0}$, or as the pairing part of the Gogny force~\cite{Berger84},
 \beq\begin{split}\label{Gogny}
V(\svec r, \svec r') = &\sum_{i = 1, 2}e^{\lrb{\lrb{r-r'}/\mu_i}^2}\\&~~~\times\lrb{W_i +
B_iP^\sigma - H_i P^\tau - M_iP^\sigma P^\tau},
 \end{split}\eeq
with the parameters $\mu_{i}$, $W_{i}$, $B_{i}$, $H_{i}$ and $M_{i}$ ($i=1,2$).

In spherically symmetric systems the solution of the RHFB equations, i.e., the Dirac spinor $\psi_{U_\alpha}$ and $\psi_{V_\alpha}$ can be written {similarly to} Eq. (\ref{Spinor}),
\begin{align}
\psi_{U_{\alpha}}({\mbox{\boldmath${ r}$}})=  &  \frac{1}{r}\left(
\begin{array}
[c]{c}%
iG_{U_{a}}(r){\mathcal{Y}}_{j_{a}m_{a}}^{l_{a}}(\hat{{\mbox{\boldmath${ r}$}}%
})\\[0.5em]
-F_{U_{a}}(r){\mathcal{Y}}_{j_{a}m_{a}}^{l_{a}^{\prime}}(\hat
{{\mbox{\boldmath${ r}$}}})
\end{array}
\right)  ,\label{SpinorUV}\\
\psi_{V_{\alpha}}({\mbox{\boldmath${ r}$}})=  &  \frac{1}{r}\left(
\begin{array}
[c]{c}
iG_{V_{a}}(r){\mathcal{Y}}_{j_{a}m_{a}}^{l_{a}}(\hat{{\mbox{\boldmath${ r}$}}%
})\\[0.5em]
-F_{V_{a}}(r){\mathcal{Y}}_{j_{a}m_{a}}^{l_{a}^{\prime}}(\hat {{\mbox{\boldmath${ r}$}}})
\end{array}
\right)  .
\end{align}
The RHFB equations (\ref{RHFBEQ}) are then reduced to the system of coupled integro-differential equations,
 \bsub\label{RHFB-R}\begin{align}
&\lrs{\frac{d}{dr} + \frac{\kappa_a}{r} + \Sigma_T} G_{U_a}(r) - \lrb{E_a + \lambda - \Sigma_-}
F_{U_a}(r)\re&~~~~~~~~~~ + X_{U_a} (r) + r\int r'dr' \Delta_a(r, r') F_{V_a}(r') = 0,\\
&\lrs{\frac{d}{dr} - \frac{\kappa_a}{r} - \Sigma_T} F_{U_a}(r) + \lrb{E_a + \lambda - \Sigma_+}
G_{U_a}(r)\re&~~~~~~~~~~ - Y_{U_a} (r) + r\int r'dr' \Delta_a(r, r') G_{V_a}(r') = 0,\\
&\lrs{\frac{d}{dr} + \frac{\kappa_a}{r} + \Sigma_T} G_{V_a}(r) + \lrb{E_a - \lambda + \Sigma_-}
F_{V_a}(r)\re&~~~~~~~~~~ + X_{V_a} (r) + r\int r'dr' \Delta_a(r, r') F_{U_a}(r') = 0,\\
&\lrs{\frac{d}{dr} - \frac{\kappa_a}{r} - \Sigma_T} F_{V_a}(r) -
\lrb{E_a - \lambda + \Sigma_+} G_{V_a}(r)\re&~~~~~~~~~~ - Y_{V_a}
(r) + r\int r'dr' \Delta_a(r, r') G_{U_a}(r') = 0,
 \end{align}\esub
where $E_{a}$ are the quasi-particle energies (without the rest mass), and the local self-energies $\Sigma_{+}$ and $\Sigma_{-}$ are
 \begin{align}
\Sigma_{+}\equiv&\Sigma_{0}+\Sigma_{S},&\Sigma_{-}\equiv&\Sigma_{0}-\Sigma _{S}-2M.
 \end{align}
In the radial RHFB equations (\ref{RHFB-R}), $X_{U}$, $Y_{U}$, $X_{V}$ and $Y_{V}$ denote the contributions from the Fock terms, which are of {a general form similar to Eq. (\ref{XY})},
 \bsub\label{Fock-terms}\begin{align}
\lrb{\bea{c}X_{U_a}\\[0.5em]Y_{U_a}\eea}_r = \int dr' \lrb{\bea{cc}X_{G_a}& X_{F_a}\\[0.5em]
Y_{G_a}& Y_{F_a}\eea}_{(r,r')} & \lrb{\bea{c} G_{U_a} \\[0.5em] F_{U_a}\eea}_{r'},\\
\lrb{\bea{c}X_{V_a}\\[0.5em]Y_{V_a}\eea}_r = \int dr' \lrb{\bea{cc}X_{G_a}& X_{F_a}\\[0.5em]
Y_{G_a}& Y_{F_a}\eea}_{(r,r')} & \lrb{\bea{c} G_{V_a} \\[0.5em] F_{V_a}\eea}_{r'}.
 \end{align}\esub
For the nonlocal terms $X_{G}$, $X_{F}$, $Y_{G}$ and $Y_{F}$ above, one needs to replace the $G$ and $F$ components in Eqs. (\ref{Non-local}) by the corresponding $G_{V}$ and $F_{V}$ in the general case{, or by $G_U$ and $F_U$ in the case of blocking}.

The pairing potentials $\Delta_a(r,r^{\prime})$ in Eqs. (\ref{RHFB-R}) can be expressed as
\begin{equation}
\Delta_{a}(r,r^{\prime})=-\sum_{b}V_{ab}^{pp}(r,r^{\prime})\kappa
_{b}(r,r^{\prime}),
\end{equation}
where the pairing tensor $\kappa(r,r^{\prime})$ reads as
 \beq\begin{split}
\kappa_b(r,r') =& \ff2\hat j_b^2\lrs{G_{U_b}(r) G_{V_b}(r') + F_{U_b}(r) F_{V_b}(r')}\\ + &\ff2\hat
j_b^2\lrs{G_{V_b}(r) G_{U_b}(r') + F_{V_b}(r) F_{U_b}(r')}.
 \end{split}\eeq
Details of the pairing interaction matrix element $V_{ab}^{pp}$ can be found in Ref.~\cite{Meng:1998a}.

\subsection{{RHFB equations in Dirac Woods-Saxon basis}}\label{sec:NUM}

In contrast to the RHB approach with $\delta$-forces in the pairing channel where the radial equations (\ref{RHFB-R}) {become} differential equations, in RHFB theory the radial equations are {fully} integro-differential. For zero-range $\delta$-forces in the pairing channel the integral terms arise from the Fock terms, and for finite-range pairing forces {they also come} from the pairing channel. {In coordinate space, it is difficult to solve such equations e.g., by a localization procedure similar to that adopted in Refs.~\cite{Bouyssy:1987, Long:2006}.} We therefore {choose} to solve them by an expansion of the Dirac-Bogoliubov spinors in an appropriate basis.

In this work {we solve the radial RHFB equations (\ref{RHFB-R}) by using the Dirac Woods-Saxon (DWS) basis introduced by Zhou et al.~\cite{Zhou:2003}.} This basis has been constructed for the investigation of weakly-bound nuclei. The set of DWS basis functions
\begin{equation}
\left\{  \left[  \varepsilon_{b},g_{\beta}({\mbox{\boldmath${ r}$}},\tau)\right]  ;\varepsilon_{b}\gtrless0\right\}  ,
\label{WSbase}%
\end{equation}
are eigenfunctions (with eigenvalues $\varepsilon_{b}$) of a Dirac equation with Woods-Saxon-like potentials for $\Sigma_{0}(r)\pm\Sigma_{S}(r)$. They are determined by the shooting method in {coordinate} space within a spherical box of size $R_{\text{max}}$~\cite{Koepf:1991}.

The $U$ and $V$ components of the Dirac Bogoliubov spinors (\ref{SpinorUV}) can be expanded as,
 \bsub\label{WS-Expand}\begin{align}
\psi_{U}=&\sum_{p=1}^{N_{F}}U_{p}g_{p}+\sum_{d=1}^{N_{D}}U_{d}g_{d},\\
\psi_{V}=&\sum_{p=1}^{N_{F}}V_{p}g_{p}+\sum_{d=1}^{N_{D}}V_{d}g_{d},
 \end{align}\esub
where $N_{F}$ and $N_{D}$ respectively correspond to the numbers of positive ($\varepsilon_{p}>0$) and negative ($\varepsilon_{d}<0$) energy states in the DWS basis. Obviously, because of spherical symmetry the quantum number $\kappa$ is preserved, i.e., the RHFB equations have to be solved for each value of $\kappa$ and the sums in the expansion (\ref{WS-Expand}) run only over states with the same $\kappa$.  For a fixed value of $\kappa$ we have the radial basis spinors
\begin{align}
g_{p}({r})=&\left(\begin{array}[c]{c}G_{p}(r)\\[0.5em]
F_{p}(r)\end{array} \right),&
g_{d}({r})=&\left(\begin{array}[c]{c}G_{d}(r)\\[0.5em]
F_{d}(r)\end{array} \right),%
\label{DWS-basis}
\end{align}
where the sub-indices $p$ and $d$ correspond to the number of nodes of the basis functions $G_{p}$ for positive energy and $F_{d}$ for negative energy.

In the DWS basis (\ref{WS-Expand}) the radial RHFB equations (\ref{RHFB-R}) are transformed to a matrix eigenvalue problem,
 \beq\label{eigenUV}
\lrb{\bea{cc}H-\lambda & \Delta \\[0.5em] \Delta & -H+\lambda\eea}\lrb{\bea{c}U\\[0.5em] V\eea}
E\lrb{\bea{c}U\\[0.5em] V\eea},
 \eeq
where $H$ and $\Delta$ are $\left(  N_{F}+N_{D}\right) \times\left( N_{F}+N_{D}\right)  $ dimensional matrices, $U$ and $V$ are the column vectors with $\left(  N_{F}+N_{D}\right)  $ elements. From the expressions of the single-particle Hamiltonian $h$ and pairing potential $\Delta$ given in the previous part we obtain the matrix elements of $H$ and $\Delta$ as
 \bsub\begin{align}
H_{nn'}^{\text{kin}} = \int dr&  G_{n}\lrb{-\frac{d}{dr}+\frac{\kappa}{r}} F_{n'}\re & + \int dr
F_{n}\lrb{\frac{d}{dr}+ \frac{\kappa}{r}}G_{n'},\\
H_{nn'}^{\text{D}} = \int dr & \lrs{ G_{n}G_{n'}\Sigma_+ + F_{n}F_{n'}\Sigma_- }\re & + \int dr {
\lrb{G_{n}F_{n'} + G_{n}F_{n'}}\Sigma_T},\\
H^{\text{E}}_{nn'} = \int dr & \int dr'\lrb{\bea{cc}G_{n}& F_{n}\eea}_r\re &\times\lrb{\bea{cc}
Y_{G} &Y_{F}\\[0.5em]X_{G}&X_{F}\eea}_{(r,r')}\lrb{\bea{c}G_{n'}\\[0.5em] F_{n'}\eea}_{r'},\\
\Delta_{nn'}= \int dr & \int dr'\Delta_\kappa(r,r') \re &\times\lrs{ G_{n}(r)G_{n'}(r') +
F_{n}(r)F_{n'}(r')},
 \end{align}\esub
where $n,n^{\prime}$ run over the radial quantum numbers of the DWS basis states in Eq. (\ref{DWS-basis}) with both positive energies ($n, n'=p$) and negative energies ($n, n'=d$).

Before carrying out RHFB applications with the DWS basis, two constituents should be firstly decided, i.e., the size of the spherical box $R_{\text{max}}$ and the number of states ($N_{F}$ and $N_{D}$) involved in the expansions (\ref{WS-Expand}). In practice, it is accurate enough to adopt the parameters of the DWS basis as $R_{\max}$ = 20fm, $N_{F}$ = 28, $N_{D}$ = 12 for the general applications whereas for weakly bound nuclear systems one needs to choose a larger spherical box radius ($R_{\max}=$ 24fm) and a larger number of states ($N_F=$36).

\section{General applications of the RHFB theory}
\label{sec:Results}

We firstly examine the equivalence between different pairing mechanisms {for stable} nuclear systems. By using the parameter set PKA1~\cite{Long:2007}, we perform the calculations for the even-even Sn isotopes from $^{106}$Sn to $^{136}$Sn by RHFB theory with Gogny and Delta pairing forces (referred to respectively by Gogny and Delta), and by DDRHF with BCS pairing (denoted by BCS($\delta$))~\cite{Long:2007}. The comparisons are based on the fact that equivalent pairing gaps are obtained with different pairing treatments. For the DDRHF calculation with BCS pairing, it is performed completely in coordinate space~\cite{Long:2007}.

\begin{table}[htbp]
\caption{Binding energy $E_B/A$ and neutron radii $r_n$ for even-even Sn isotopes. The results are calculated by RHFB with Gogny and Delta pairing forces, and by DDRHF with BCS pairing~\cite{Long:2007}, in comparison with the experimental data~\cite{Audi:2003}. The used parameter set is PKA1~\cite{Long:2007}. }\label{tab:Sn-E}
\begin{tabular}{c|cccc|ccc}\hline\hline
         &\multicolumn{4}{c|}{$E/A$ (MeV)}       &\multicolumn{3}{c}{$r_n$(fm)} \\
 $N$     &Exp      &  Gogny  &   Delta & BCS($\delta$)&Gogny   &Delta   &BCS($\delta$)\\ \hline
 56      &$-$8.4327&$-$8.4339&$-$8.4423&$-$8.4425&4.456   &4.451   &4.470  \\
 58      &$-$8.4688&$-$8.4605&$-$8.4687&$-$8.4694&4.508   &4.501   &4.523  \\
 60      &$-$8.4961&$-$8.4804&$-$8.4877&$-$8.4889&4.558   &4.550   &4.573  \\
 62      &$-$8.5137&$-$8.4940&$-$8.5000&$-$8.5017&4.606   &4.597   &4.620  \\
 64      &$-$8.5226&$-$8.5018&$-$8.5063&$-$8.5085&4.651   &4.642   &4.665  \\
 66      &$-$8.5231&$-$8.5039&$-$8.5071&$-$8.5097&4.695   &4.686   &4.708  \\
 68      &$-$8.5166&$-$8.5006&$-$8.5029&$-$8.5058&4.735   &4.728   &4.748  \\
 70      &$-$8.5045&$-$8.4921&$-$8.4937&$-$8.4969&4.772   &4.767   &4.785  \\
 72      &$-$8.4879&$-$8.4788&$-$8.4799&$-$8.4833&4.805   &4.802   &4.818  \\
 74      &$-$8.4674&$-$8.4613&$-$8.4616&$-$8.4652&4.835   &4.834   &4.847  \\
 76      &$-$8.4436&$-$8.4401&$-$8.4396&$-$8.4431&4.863   &4.863   &4.874  \\
 78      &$-$8.4168&$-$8.4157&$-$8.4145&$-$8.4175&4.889   &4.889   &4.897  \\
 80      &$-$8.3869&$-$8.3882&$-$8.3871&$-$8.3889&4.913   &4.913   &4.917  \\\hline
 82      &$-$8.3549&$-$8.3579&$-$8.3579&$-$8.3579&4.935   &4.935   &4.935  \\\hline
 84      &$-$8.2779&$-$8.2752&$-$8.2744&$-$8.2733&4.993   &4.991   &5.001  \\
 86      &$-$8.1990&$-$8.1934&$-$8.1916&$-$8.1900&5.050   &5.046   &5.062  \\  
 \hline\hline
\end{tabular}
\end{table}

In Table \ref{tab:Sn-E} are shown the binding energy $E_B/A$ and neutron radii $r_n$, extracted from the calculations with Bogoliubov and BCS pairings. From Table \ref{tab:Sn-E} one can find good agreement on the binding energies since the studied nuclei are located in the stability valley. For the neutron radii, there exist some minor systematic deviations between the results of Bogoliubov and BCS pairings. Except for the magic nuclei, the calculations with
BCS pairing present slightly larger values ($\sim0.01$fm) than those given by Bogoliubov pairings.

\begin{figure}[htbp]
\includegraphics[width = 0.47\textwidth]{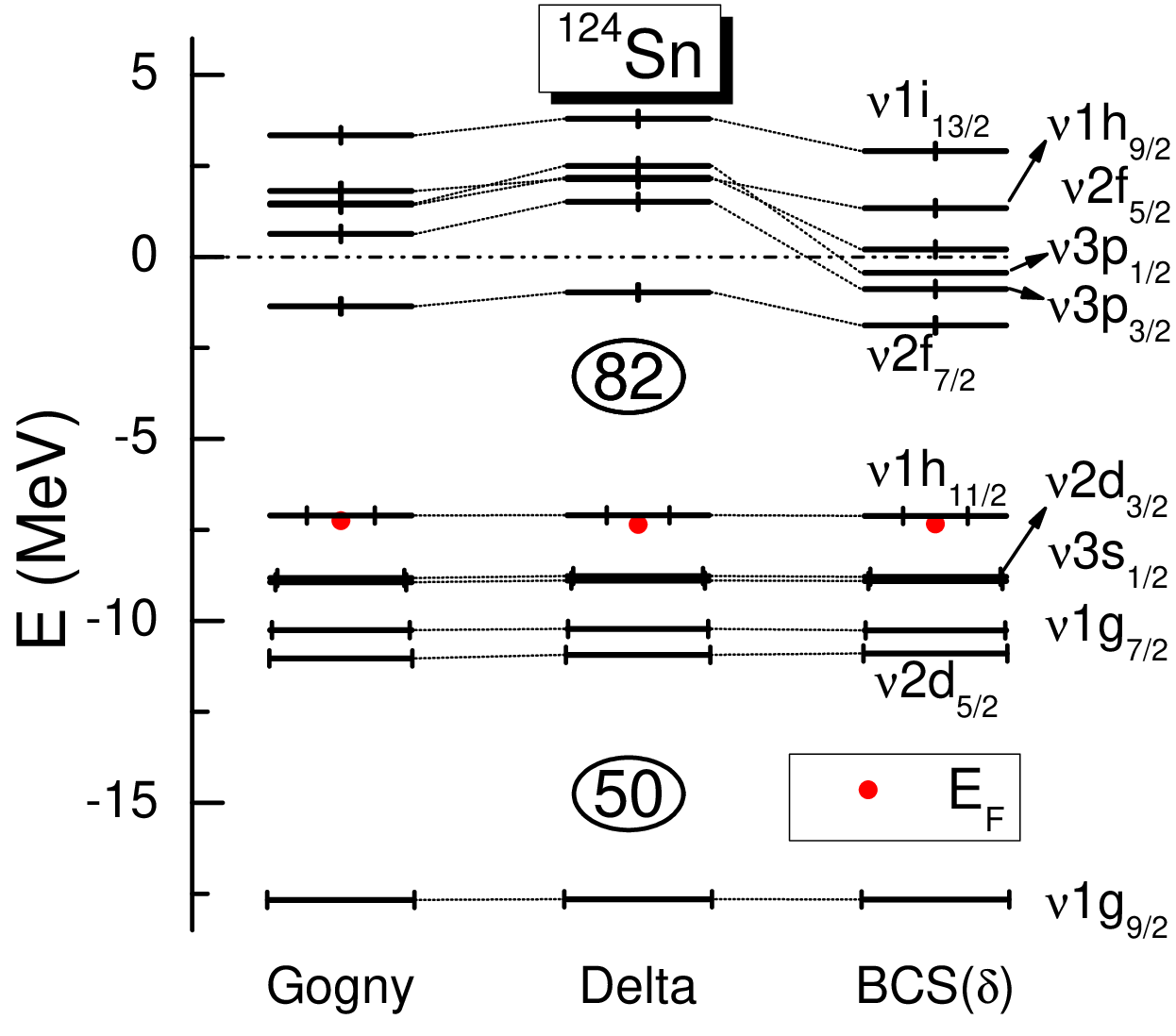}
\caption{Neutron canonical single-particle energies for $^{124}$Sn,
calculated by RHFB with Gogny and Delta pairing forces, and by DDRHF
with BCS pairing~\cite{Long:2007}. Horizontal error bars denote the
occupation probabilities of the states and filled circles represent
the Fermi levels. See the text for details. }\label{fig:Sn124LevN}
\end{figure}

Taking $^{124}$Sn as an example, in Fig. \ref{fig:Sn124LevN} are shown the neutron canonical single-particle energies and the occupation probabilities (in horizontal error bars) extracted from RHFB calculations with Gogny and Delta pairing forces. For comparison are also shown the results from DDRHF calculations with BCS pairing. In the Bogoliubov scheme the canonical single-particle states, i.e., the eigenstates of the density matrix, can be obtained with the
canonical transformation from the Bogoliubov quasi-particle to the canonical basis~\cite{Ring:1980}. With the BCS approximation the density matrix and single-particle Hamiltonian do commute. The corresponding single-particle energies are therefore the canonical ones.

\begin{figure*}[htpb]
\includegraphics[width = 0.8\textwidth]{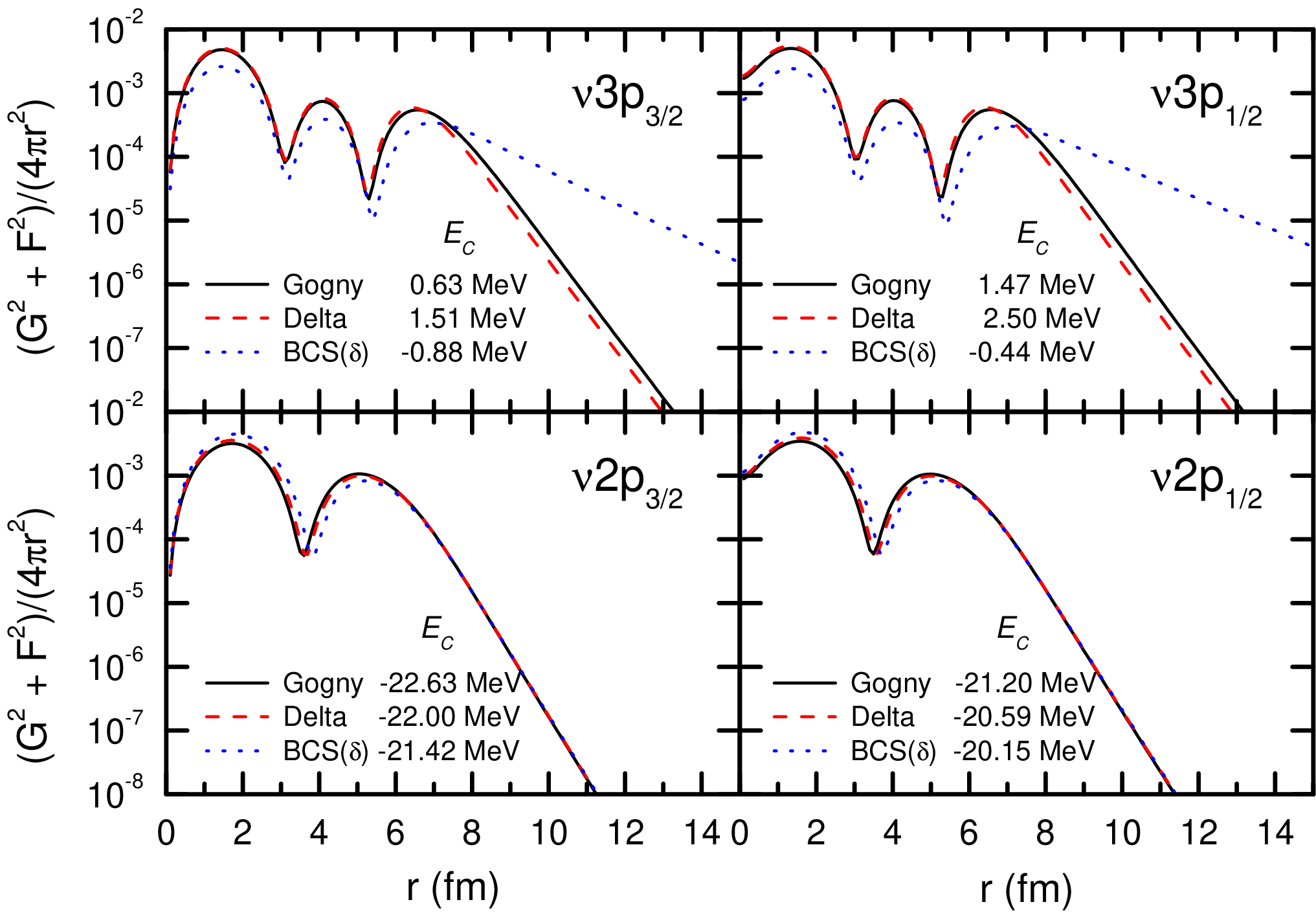}
\caption{The occupation densities of the $\nu2p$ and $\nu3p$ states, extracted from RHFB calculation with Gogny and Delta pairing forces, and from DDRHF (BCS($\delta$)) with BCS pairing~\cite{Long:2007}. See the text for details. } \label{fig:Sn124State}
\end{figure*}

As shown in Fig. \ref{fig:Sn124LevN} there is no distinct difference in the occupation probabilities  (denoted by horizontal error bars) between different pairing treatments because of the existence of the shell gap $82$. For the single-particle energies, the calculations with Bogoliubov and BCS pairings provide identical values for the states below the Fermi level. For the states above, particulary the low-$l$ ones, remarkable deviations are found. As seen from the occupation densities in Fig. \ref{fig:Sn124State}, different pairing treatments lead to identical radial  distributions for the deeply bound $\nu2p$ states. For the $\nu3p$ states, the occupation densities given by BCS calculations become rather diffuse at large distance although they are weakly bound. In contrast the calculations
with Bogoliubov pairings still present appropriate asymptotic behavior at large distance even when the states lie beyond the particle continuum threshold. From Fig. \ref{fig:Sn124State} one may recognize that within the Bogoliubov scheme the occupation densities are properly localized inside the nucleus such that the continuum effects can be efficiently taken into account. For the stable nuclei, this is less important, e.g., in $^{124}$Sn the scattering of Cooper pairs into the continuum is blocked by the shell gap $82$. In the weakly bound nuclei the valence orbits may gather around the  particle continuum threshold and the continuum effects are then strongly enhanced. As shown in Fig. \ref{fig:Sn124State}, such effects can be self-consistently and efficiently taken into account by the Bogoliubov transformation~\cite{Meng:1996, Meng:1998a}.

In the applications of the RHFB theory with the zero-range pairing force, the cut-off on the quasi-particle energy is an important ingredient as well as the pairing strength $V_{0}$. In the above calculations the pairing strength is set to $V_0=325$MeV with the quasi-particle energy cut-off $\sim100MeV$. Compared to the zero-range pairing force, the finite range Gogny force is of less arbitrariness because of the finite range and natural cut-off. In addition, an appropriate description of the mean field can also be provided by the Gogny force in the non-relativistic calculations and therefore better systematics is expected with the Gogny-type pairing force.

Now we aim for the systematical study of both pairing correlations and mean fields by considering Sn isotopes from $^{100}$Sn to $^{137}$Sn, and $N=82$ isotones from $^{129}$Ag to $^{153}$Lu as representatives. The calculations use the RHFB theory with the parameter sets PKA1~\cite{Long:2007} (with $\rho$-tensor couplings) and PKO1~\cite{Long:2006} (without $\rho$-tensor couplings), and they are compared to those obtained by RHB theory with the parameter set DD-ME2~\cite{Lalazissis:2005}, one of the {most successful} candidates in the existing RMF effective interactions. In the following, the finite range Gogny force D1S~\cite{Berger84} is adopted in pairing channel. For the isotopes (isotones) with {an} odd neutron (proton) number, the blocking effects have to be taken into account. In the corresponding calculations, we blocked different orbits around the Fermi surface, which can be provided by the calculations of the neighboring even isotopes or isotones, and {we} chose the state with the largest binding energy $\left\vert E_{B}\right\vert $ as the ground state.

\begin{table*}[htbp]
\caption{The binding energies per particle $E_{B}/A$ (MeV) of Sn isotopes and the blocked neutron ($\nu$) orbits $j_{b}$ of the odd isotopes. The results are calculated by RHFB with PKA1~\cite{Long:2007} and PKO1~\cite{Long:2006}, RHB with DD-ME2~\cite{Lalazissis:2005}, in comparison to the data~\cite{Audi:2003}. {The quantities $\Delta$ are the r.m.s. deviations.} } \label{tab:Sn} \setlength{\tabcolsep}{4pt}
\begin{tabular}[c]{c|c|ccc||c|c|cc|cc|cc}\hline\hline
    &  Exp    &  PKA1   &  PKO1  &  DD-ME2&    &  Exp   & \multicolumn{2}{c|}{PKA1}&\multicolumn{2}{c|}{PKO1}&\multicolumn{2}{c}{DD-ME2}\\
 $N$&  $E_B/A$& $E_B/A$ & $E_B/A$& $E_B/A$& $N$& $E_B/A$& $E_B/A$ &$j_b$           &$E_B/A$ &$j_b$           &$E_B/A$ &$j_b$            \\ \hline
 50 &$-$8.2479&$-$8.3097&$-$8.2831&$-$8.2635& 51 &$-$8.2740&$-$8.3242 &$\nu 2d_{5/2}$  &$-$8.3027 &$\nu 1g_{7/2}$  &$-$8.2804 &$\nu 1g_{7/2}$   \\
 52 &$-$8.3244&$-$8.3587&$-$8.3454&$-$8.3198& 53 &$-$8.3420&$-$8.3686 &$\nu 2d_{5/2}$  &$-$8.3588 &$\nu 1g_{7/2}$  &$-$8.3327 &$\nu 1g_{7/2}$   \\
 54 &$-$8.3836&$-$8.4001&$-$8.3969&$-$8.3688& 55 &$-$8.3965&$-$8.4047 &$\nu 2d_{5/2}$  &$-$8.4046 &$\nu 1g_{7/2}$  &$-$8.3778 &$\nu 1g_{7/2}$   \\
 56 &$-$8.4327&$-$8.4340&$-$8.4390&$-$8.4109& 57 &$-$8.4401&$-$8.4327 &$\nu 2d_{5/2}$  &$-$8.4413 &$\nu 2d_{5/2}$  &$-$8.4158 &$\nu 2d_{5/2}$   \\
 58 &$-$8.4688&$-$8.4606&$-$8.4724&$-$8.4463& 59 &$-$8.4706&$-$8.4551 &$\nu 1g_{7/2}$  &$-$8.4715 &$\nu 2d_{5/2}$  &$-$8.4487 &$\nu 2d_{5/2}$   \\
 60 &$-$8.4961&$-$8.4805&$-$8.4977&$-$8.4746& 61 &$-$8.4932&$-$8.4733 &$\nu 3s_{1/2}$  &$-$8.4928 &$\nu 2d_{5/2}$  &$-$8.4740 &$\nu 2d_{5/2}$   \\
 62 &$-$8.5137&$-$8.4942&$-$8.5149&$-$8.4957& 63 &$-$8.5069&$-$8.4854 &$\nu 3s_{1/2}$  &$-$8.5049 &$\nu 2d_{5/2}$  &$-$8.4898 &$\nu 2d_{5/2}$   \\
 64 &$-$8.5226&$-$8.5019&$-$8.5243&$-$8.5085& 65 &$-$8.5141&$-$8.4912 &$\nu 3s_{1/2}$  &$-$8.5117 &$\nu 3s_{1/2}$  &$-$8.4997 &$\nu 3s_{1/2}$   \\
 66 &$-$8.5231&$-$8.5041&$-$8.5260&$-$8.5122& 67 &$-$8.5096&$-$8.4907 &$\nu 3s_{1/2}$  &$-$8.5117 &$\nu 3s_{1/2}$  &$-$8.5010 &$\nu 3s_{1/2}$   \\
 68 &$-$8.5166&$-$8.5007&$-$8.5213&$-$8.5080& 69 &$-$8.4995&$-$8.4844 &$\nu 3s_{1/2}$  &$-$8.5046 &$\nu 3s_{1/2}$  &$-$8.4935 &$\nu 3s_{1/2}$   \\
 70 &$-$8.5045&$-$8.4922&$-$8.5110&$-$8.4976& 71 &$-$8.4853&$-$8.4725 &$\nu 3s_{1/2}$  &$-$8.4920 &$\nu 3s_{1/2}$  &$-$8.4793 &$\nu 3s_{1/2}$   \\
 72 &$-$8.4879&$-$8.4790&$-$8.4960&$-$8.4820& 73 &$-$8.4673&$-$8.4574 &$\nu 1h_{11/2}$ &$-$8.4744 &$\nu 3s_{1/2}$  &$-$8.4610 &$\nu 1h_{11/2}$  \\
 74 &$-$8.4674&$-$8.4615&$-$8.4768&$-$8.4624& 75 &$-$8.4456&$-$8.4391 &$\nu 1h_{11/2}$ &$-$8.4529 &$\nu 1h_{11/2}$ &$-$8.4406 &$\nu 1h_{11/2}$  \\
 76 &$-$8.4436&$-$8.4403&$-$8.4536&$-$8.4395& 77 &$-$8.4208&$-$8.4171 &$\nu 1h_{11/2}$ &$-$8.4285 &$\nu 1h_{11/2}$ &$-$8.4169 &$\nu 1h_{11/2}$  \\
 78 &$-$8.4168&$-$8.4158&$-$8.4265&$-$8.4139& 79 &$-$8.3928&$-$8.3917 &$\nu 1h_{11/2}$ &$-$8.4002 &$\nu 1h_{11/2}$ &$-$8.3905 &$\nu 1h_{11/2}$  \\
 80 &$-$8.3869&$-$8.3883&$-$8.3956&$-$8.3858& 81 &$-$8.3629&$-$8.3633 &$\nu 1h_{11/2}$ &$-$8.3677 &$\nu 1h_{11/2}$ &$-$8.3618 &$\nu 1h_{11/2}$  \\
 82 &$-$8.3549&$-$8.3580&$-$8.3605&$-$8.3556& 83 &$-$8.3107&$-$8.3103 &$\nu 2f_{7/2}$  &$-$8.3093 &$\nu 2f_{7/2}$  &$-$8.3034 &$\nu 2f_{7/2}$   \\
 84 &$-$8.2779&$-$8.2754&$-$8.2757&$-$8.2644& 85 &$-$8.2320&$-$8.2277 &$\nu 2f_{7/2}$  &$-$8.2246 &$\nu 2f_{7/2}$  &$-$8.2123 &$\nu 2f_{7/2}$   \\
 86 &$-$8.1990&$-$8.1936&$-$8.1921&$-$8.1744& 87 &$-$8.1530&$-$8.1455 &$\nu 2f_{7/2}$  &$-$8.1413 &$\nu 2f_{7/2}$  &$-$8.1222 &$\nu 2f_{7/2}$   \\\hline
 $\Delta$&   &  0.0197& 0.0115 &0.0137  & $\Delta$&   &   0.0177&                &0.0095  &                &0.0146  &                 \\
\hline\hline
\end{tabular}
\end{table*}

\begin{table*}[htbp]
\caption{Same as {Table \ref{tab:Sn}}, but for $N=82$ isotones}%
\label{tab:N82}
\setlength{\tabcolsep}{3pt}
\begin{tabular}
[c]{l|c|ccc||l|c|cc|cc|cc}\hline\hline
          & Exp    &  PKA1  &  PKO1  &  DD-ME2&           &  Exp   & \multicolumn{2}{c|}{PKA1}&\multicolumn{2}{c|}{PKO1}&\multicolumn{2}{c}{DD-ME2}\\
          & $E_B/A$& $E_B/A$& $E_B/A$& $E_B/A$&           & $E_B/A$& $E_B/A$& $j_b$        &$E_B/A$  &$j_b$           &$E_B/A$ &$j_b$              \\ \hline
$^{130}$Cd&$-$8.2561&$-$8.2563&$-$8.2699&$-$8.2491& $^{129}$Ag&$-$8.1930&$-$8.1887 &$\pi1g_{9/2}$ &$-$8.2046& $\pi1g_{9/2}$  &$-$8.1772& $\pi1g_{9/2}$ \\
$^{132}$Sn&$-$8.3549&$-$8.3580&$-$8.3605&$-$8.3556& $^{131}$In&$-$8.2988&$-$8.2989 &$\pi1g_{9/2}$ &$-$8.3056& $\pi1g_{9/2}$  &$-$8.2941& $\pi1g_{9/2}$ \\
$^{134}$Te&$-$8.3838&$-$8.3818&$-$8.3998&$-$8.3888& $^{133}$Sb&$-$8.3649&$-$8.3626 &$\pi1g_{7/2}$ &$-$8.3730& $\pi1g_{7/2}$  &$-$8.3659& $\pi1g_{7/2}$ \\
$^{136}$Xe&$-$8.3962&$-$8.3912&$-$8.4208&$-$8.4063& $^{135}$I &$-$8.3848&$-$8.3788 &$\pi1g_{7/2}$ &$-$8.4031& $\pi1g_{7/2}$  &$-$8.3911& $\pi1g_{7/2}$ \\
$^{138}$Ba&$-$8.3934&$-$8.3869&$-$8.4253&$-$8.4089& $^{137}$Cs&$-$8.3890&$-$8.3807 &$\pi1g_{7/2}$ &$-$8.4157& $\pi1g_{7/2}$  &$-$8.4009& $\pi1g_{7/2}$ \\
$^{140}$Ce&$-$8.3764&$-$8.3694&$-$8.4123&$-$8.3956& $^{139}$La&$-$8.3781&$-$8.3685 &$\pi1g_{7/2}$ &$-$8.4107& $\pi1g_{7/2}$  &$-$8.3951& $\pi1g_{7/2}$ \\
$^{142}$Nd&$-$8.3461&$-$8.3395&$-$8.3787&$-$8.3618& $^{141}$Pr&$-$8.3540&$-$8.3453 &$\pi2d_{5/2}$ &$-$8.3869& $\pi2d_{5/2}$  &$-$8.3715& $\pi2d_{5/2}$ \\
$^{144}$Sm&$-$8.3037&$-$8.2979&$-$8.3312&$-$8.3140& $^{143}$Pm&$-$8.3178&$-$8.3097 &$\pi2d_{5/2}$ &$-$8.3464& $\pi2d_{5/2}$  &$-$8.3305& $\pi2d_{5/2}$ \\
$^{146}$Gd&$-$8.2496&$-$8.2449&$-$8.2723&$-$8.2548& $^{145}$Eu&$-$8.2693&$-$8.2613 &$\pi2d_{5/2}$ &$-$8.2922& $\pi2d_{5/2}$  &$-$8.2759& $\pi2d_{5/2}$ \\
$^{148}$Dy&$-$8.1809&$-$8.1810&$-$8.2032&$-$8.1853& $^{147}$Tb&$-$8.2067&$-$8.2022 &$\pi2d_{3/2}$ &$-$8.2268& $\pi2d_{5/2}$  &$-$8.2100& $\pi2d_{5/2}$ \\
$^{150}$Er&$-$8.1022&$-$8.1074&$-$8.1250&$-$8.1065& $^{149}$Ho&$-$8.1335&$-$8.1346 &$\pi3s_{1/2}$ &$-$8.1528& $\pi1h_{11/2}$ &$-$8.1356& $\pi1h_{11/2}$\\
$^{152}$Yb&$-$8.0157&$-$8.0252&$-$8.0384&$-$8.0196& $^{151}$Tm&$-$8.0501&$-$8.0575 &$\pi3s_{1/2}$ &$-$8.0710& $\pi1h_{11/2}$ &$-$8.0533& $\pi1h_{11/2}$\\
$^{154}$Hf&$-$7.9180&$-$7.9354&$-$7.9442&$-$7.9250& $^{153}$Lu&$-$7.9593&$-$7.9719 &$\pi3s_{1/2}$ &$-$7.9810& $\pi1h_{11/2}$ &$-$7.9629& $\pi1h_{11/2}$\\  \hline
 $\Delta$ &        &  0.0071&  0.0247&  0.0099&  $\Delta$ &        & 0.0071 &              &   0.0222&                &  0.0100&                   \\
\hline\hline
\end{tabular}
\end{table*}

\begin{figure*}[htbp]
\includegraphics[width = 0.35\textwidth]{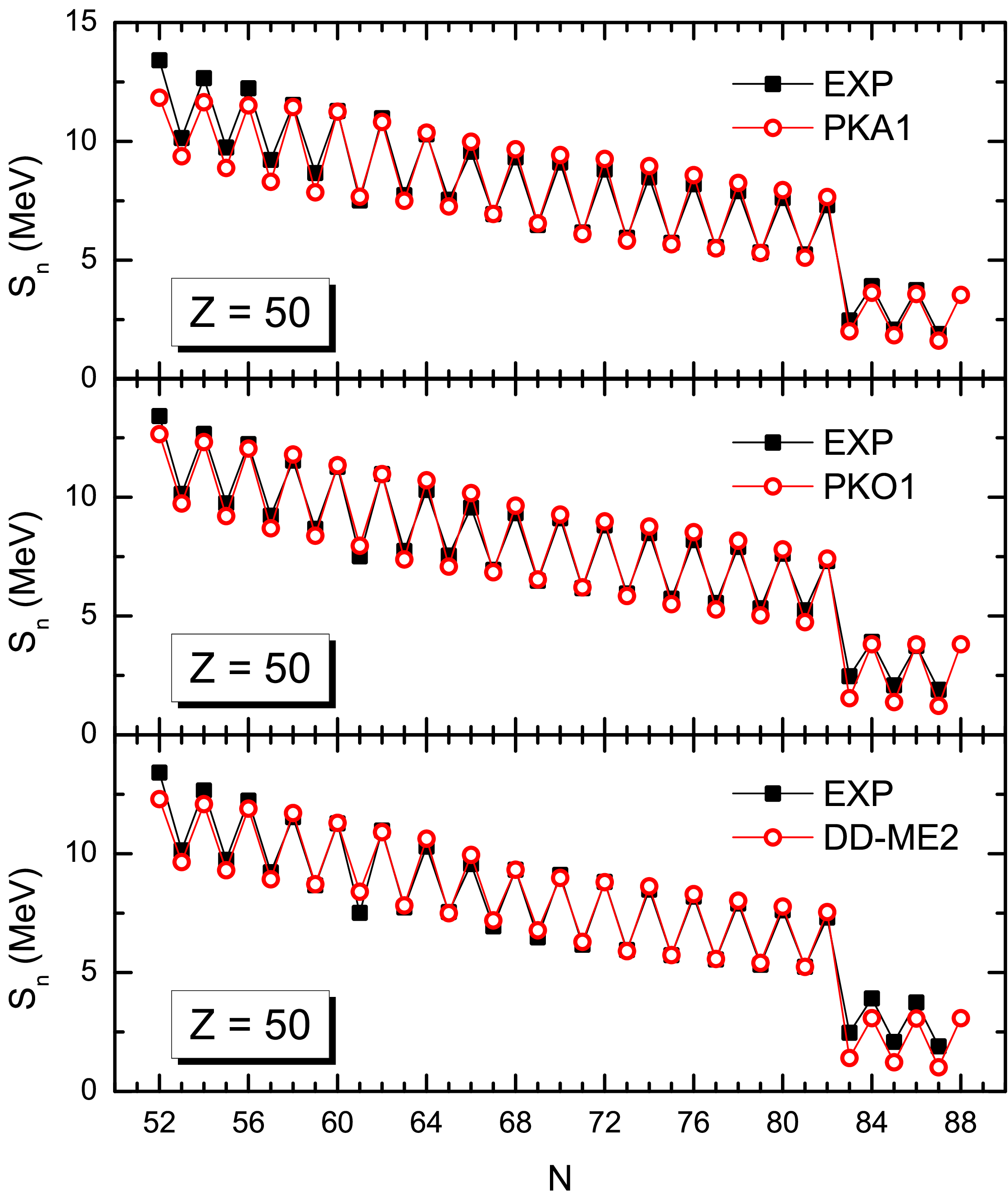}~~~
\includegraphics[width = 0.35\textwidth]{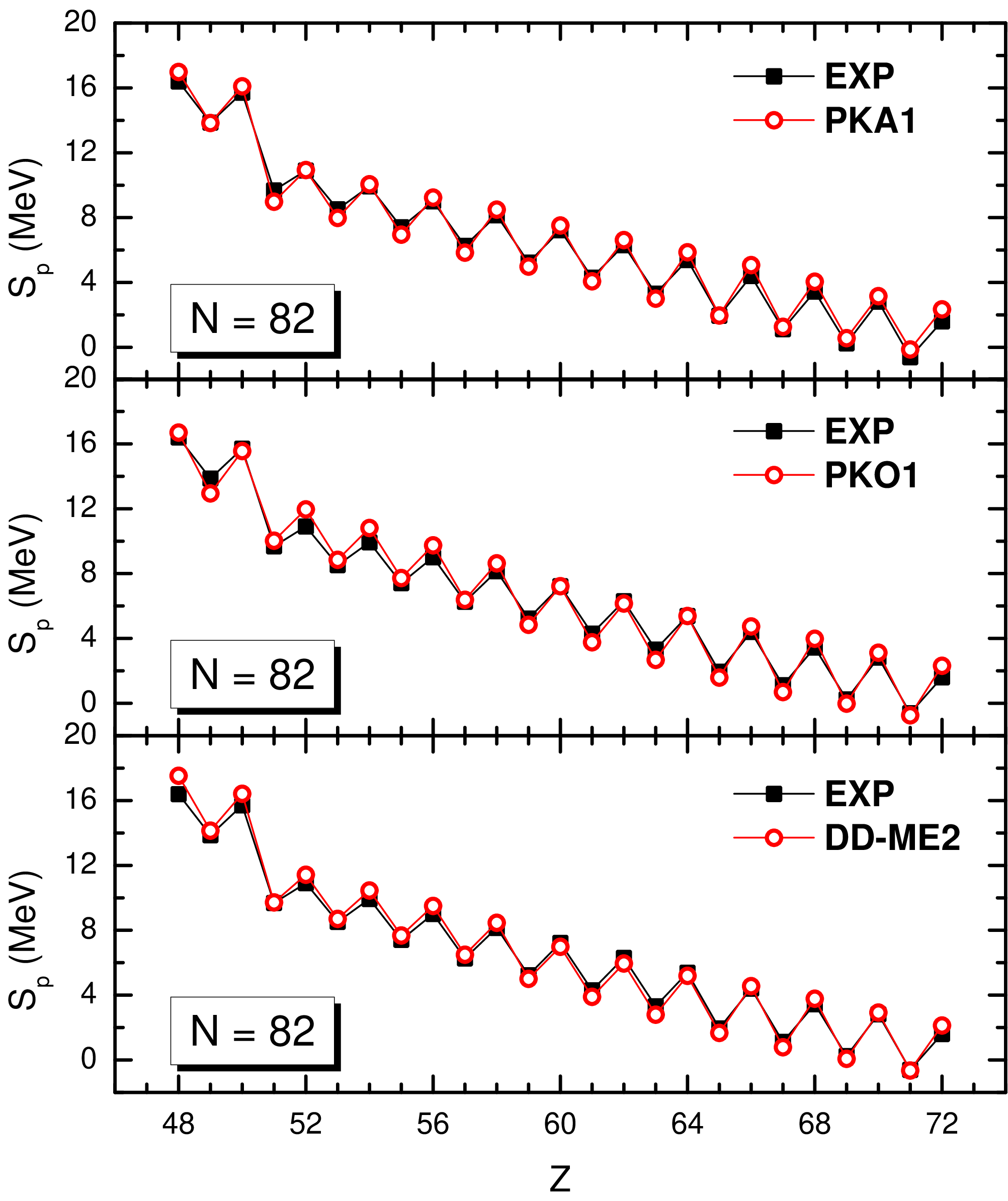}
\caption{(color online) Single-neutron separation energies (MeV) along Sn isotopic ($S_{n}$: left panels) and $N=82$ isotonic ($S_{p}$: right panels) chains. The results are calculated by RHFB with PKA1~\cite{Long:2007}, PKO1~\cite{Long:2006}, and by RHB with DD-ME2~\cite{Lalazissis:2005}, in comparison to the experimental data~\cite{Audi:2003}.}\label{fig:Sn&N82-OE}
\end{figure*}

\begin{figure*}[htbp]
\includegraphics[width = 0.35\textwidth]{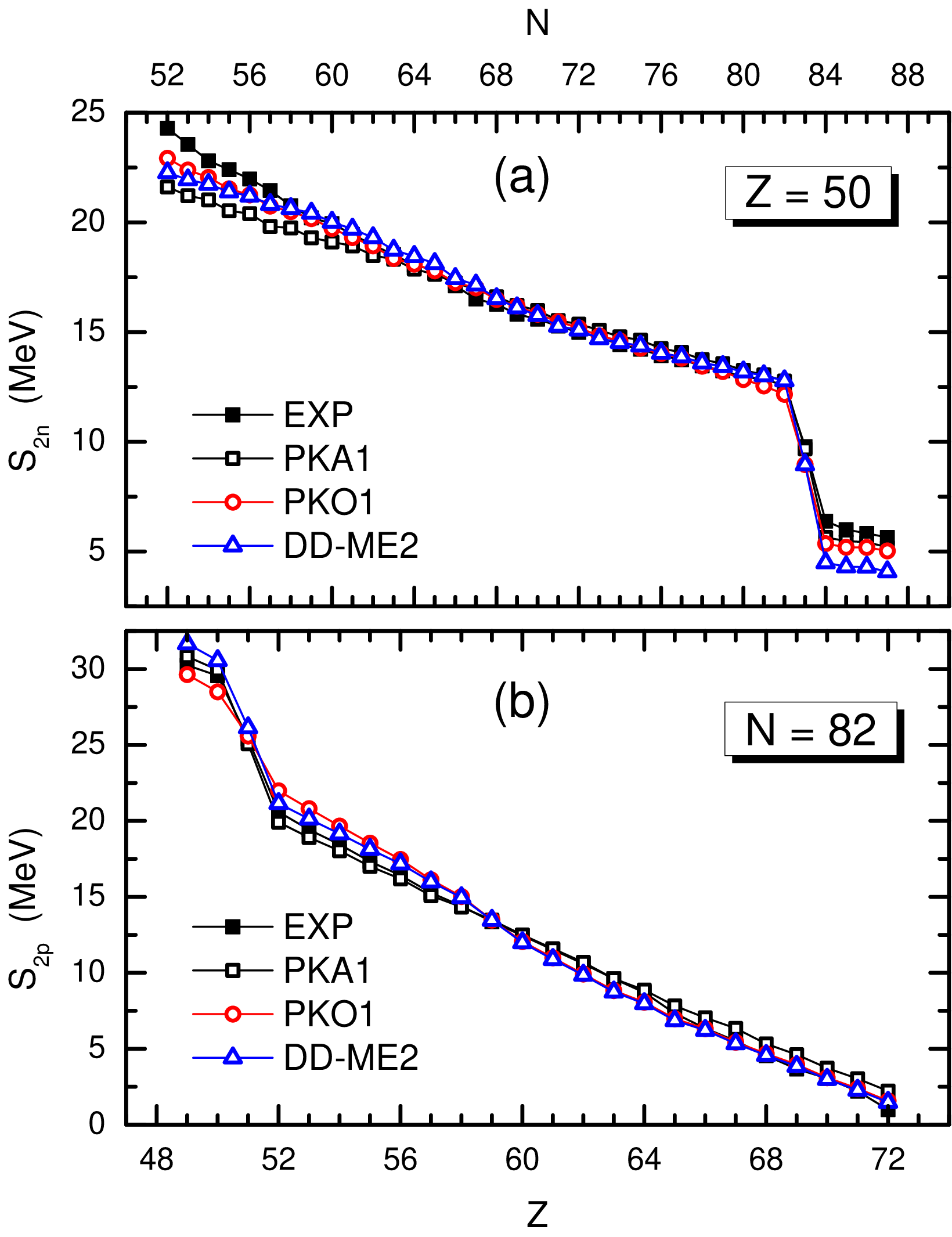}~~~
\includegraphics[width = 0.35\textwidth]{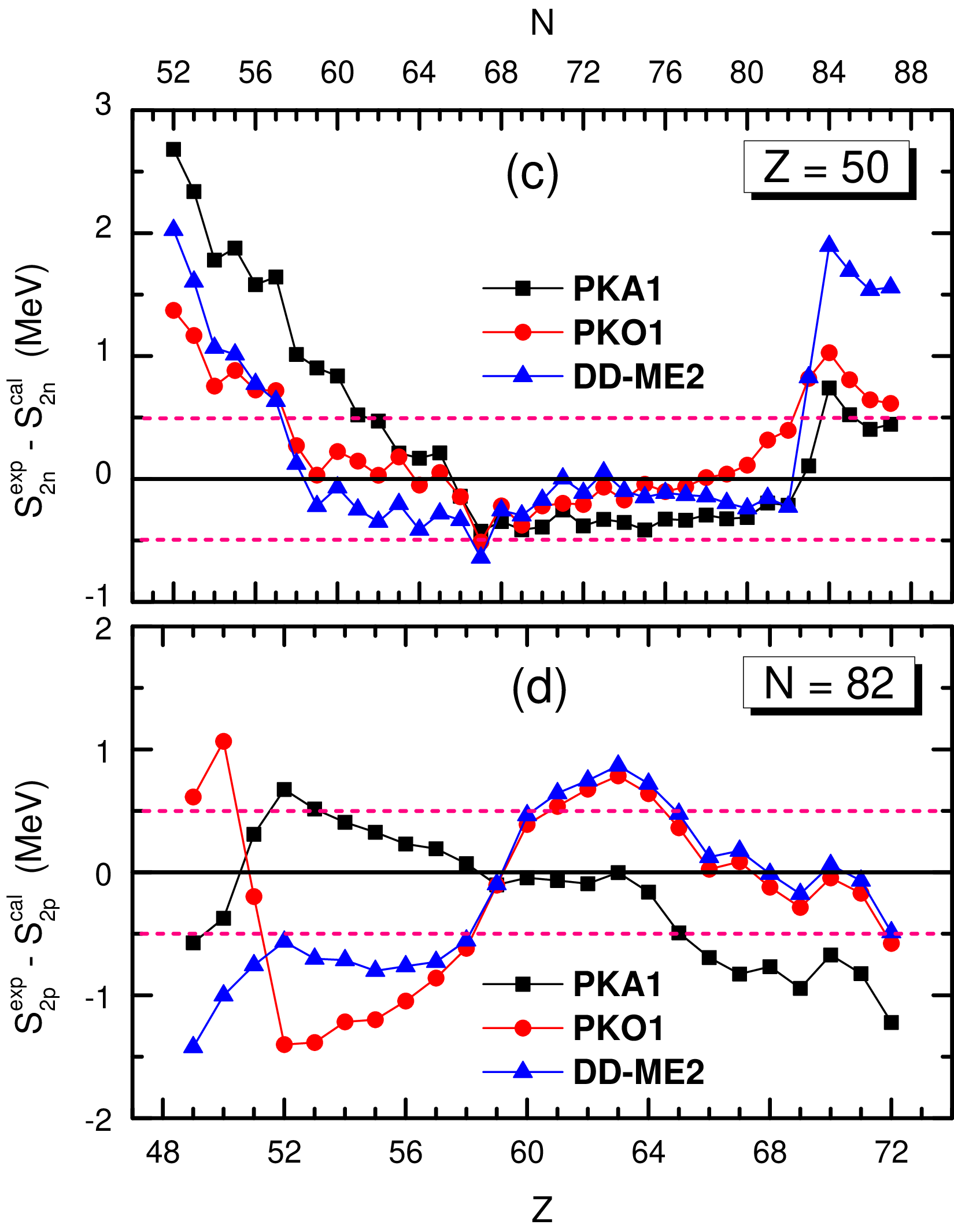}
\caption{(color online) Two-nucleon separation energies (MeV) (left panels) and the deviations (right panels) from the experimental data~\cite{Audi:2003} for Sn isotopes ($S_{2n}$: upper panel) and $N=82$ isotones ($S_{2p}$: lower panel). The results are calculated by RHFB with PKA1~\cite{Long:2007} and PKO1~\cite{Long:2006}, and by RHB with DD-ME2~\cite{Lalazissis:2005}. See the text for details.}\label{fig:S2np}
\end{figure*}

In {Table \ref{tab:Sn}} and {Table \ref{tab:N82}} we show the binding energies per particle $E_{B}/A$ for Sn isotopes and $N=82$ isotones, respectively, as well as the blocked orbits ($j_{b}$) for the odd-A isotopes. For the odd Sn isotopes we find, except for $^{123}$Sn$_{73}$, the same blocking configurations for the parameter sets PKO1 and DD-ME2, which provide similar neutron spectra, e.g., for $^{132}$Sn (see Ref.~\cite{Long:2007}). However, PKA1 shows very different blocking results for $N<65$. {This is mainly due to the fact that the pseudo-spin partners $\lrb{\nu1g_{7/2}, \nu2d_{5/2}}$ near the Fermi surface are somehow degenerate in the results of PKA1~\cite{Long:2007}. In contrast the calculations with PKO1 and DD-ME2 present remarkable gaps between these two states, i.e., the artificial shell closures $N=58$~\cite{Geng:2006, Long:2007}.} In {Table \ref{tab:Sn}} a long-range blocking is found in $\nu s_{1/2}$ (more than 4 odd isotopes), which implies that the low-$l$ states are more favored by the blocking effects. For the odd $N=82$ isotones we find in {Table \ref{tab:N82}} identical blocking on the neutron rich side ($Z\leq63$: $^{145}$Eu) for PKA1, PKO1 and DD-ME2. When $Z\geq65$ ($^{147}$Tb), PKA1 gives a different blocking, e.g., the blocking favored state $\pi1s_{1/2}$. In the last rows of {Table \ref{tab:Sn}} and {Table \ref{tab:N82}} we show the the root mean square deviations $\Delta$ (averaged over the isotopes in the column) of the binding energy $E_{B}/A$ from the experimental {values}~\cite{Audi:2003} for both even and odd nuclei. {They indicate that} the three models, RHF with $\rho$-tensor couplings (PKA1), RHF without $\rho$-tensor couplings (PKO1), and RMF (DD-ME2), {present comparable quantitative accuracies}, and PKO1 provides the best overall agreement for the Sn isotopes whereas PKA1 presents the best overall descriptions for $N=82$ isotones.

From the binding energies in {Table \ref{tab:Sn}} and {Table \ref{tab:N82}}, we {have} extracted the single-nucleon and two-nucleon separation energies to study the systematics of both mean fields and pairing correlations. {Fig. \ref{fig:Sn&N82-OE}} presents the single-neutron separation energies $S_{n}$ of Sn isotopes from $^{101}$Sn to $^{138}$Sn (left panels) and the single-proton separation energies $S_{p}$ of $N=82$ isotones from $^{130}$Cd to $^{153}$Lu (right panels), in comparison with the experimental data from Ref.~\cite{Audi:2003}. It is well known that the odd-even differences on the single-nucleon separation energies reflect the effects of the pairing correlations. In {Fig. \ref{fig:Sn&N82-OE}}, PKA1, PKO1 and DD-ME2 present comparable and {satisfactory} quantitative agreements with the data for both {isotopic and isotonic chains}, which means that the appropriate description of the pairing correlations can be provided by the RHFB theory with the finite-range Gogny pairing force. From {Fig. \ref{fig:Sn&N82-OE}}, one can find some systematics in the results of these three models. On the neutron-rich side, i.e., after $^{132}$Sn for Sn isotopes and before $_{~64}^{146}$Gd for $N=82$ isotones, PKA1 shows a better agreement than PKO1 and DD-ME2. On the proton rich side, these three models show similar accuracy.

In {Fig. \ref{fig:S2np}} are shown the two-nucleon separation energies (plot a and b) and the deviations (plot c and d) from the experimental data for Sn isotopes (plot a and c) and $N=82$ isotones (plot b and d). {It can be seen that} PKA1, PKO1 and DD-ME2 {reproduce well} the data in a rather wide range, the deviations {being} within $\pm$0.5 MeV. As we know, the sudden change on the two-nucleon separation energy in general reflects the existence of significant structure (e.g., at $^{132}$Sn). {From {Fig. \ref{fig:S2np}c and Fig. \ref{fig:S2np}d}, one can see} that PKA1 shows a different agreement from PKO1 and DD-ME2. Along the Sn isotopic chain, PKA1 presents good quantitative agreement from $N= 61$ to $87$ and large deviations are found on the proton rich side. In the results calculated by RHB with DD-ME2, {large} deviations are seen on both neutron and proton rich sides as shown in {Fig. \ref{fig:S2np}}c. Among these three effective interactions, PKO1 provides the best overall agreement with the data for Sn isotopes {while for $N=82$ isotones ({Fig. \ref{fig:S2np}d})} PKA1 presents the best overall agreement.

Concerning the separation energies, better systematics are obtained from the stable region to the neutron rich side with the inclusion of Fock terms, especially with the presence of $\rho$-tensor couplings, e.g., around $^{132}$Sn in Sn isotopic chain as well as the region around $^{140}$Ce in $N=82$ isotones (see right panels of Fig. \ref{fig:S2np}). In fact, such improvements are consistent with the elimination of the artificial shell closures 58 and 92~\cite{Geng:2006, Long:2007} beyond the magic gaps 50 and 82, which may change the mean fields and pairing effects. These artificial shell closures appear in all RMF models, and in RHF they can be eliminated with the inclusion of the $\rho$-tensor couplings. In addition, the improved systematics from the stable region to neutron rich side are meaningful for the reliable exploration of the nuclear systems with extreme neutron-to-proton ratios.

In principle, with the model Lagrangian based on meson-exchange nucleon-nucleon interactions one could have the same degrees of freedom in RMF {as in RHF}. However, the pion pseudo-vector and rho-tensor couplings cannot be efficiently taken into account by the RMF because of the lack of exchange terms. As pointed out in Refs.~\cite{Long:2007, Long:2008}, these two couplings bring indeed significant improvements on the description of the shell structure and its evolution while because of their nature, they do not bring much additional freedom to the description of binding energies. This is the reason why three different models provide equivalent accuracy on the binding energies of Sn isotopes and $N=82$ isotones. Even though, distinct deviations still exist between RHF and RMF, or between RHFB and RHB, in the systematic behaviors of the binding energies.

\vspace{0.5cm}

\section{Summary}\label{sec:Conclusions}

In this paper, we have introduced the relativistic Hartree-Fock-Bogoliubov (RHFB) theory with density-dependent meson-nucleon couplings. {The RHFB} equations are solved by an expansion of the Dirac-Bogoliubov spinors on a relativistic Dirac Woods-Saxon (DWS) basis. By taking the finite range Gogny force D1S as the pairing force, we {have} performed RHFB-DWS calculations for both stable and weakly bound nuclei. The parameters of the DWS basis are determined for the applications of {the} RHFB theory in exotic as well as stable nuclei. The quantitative agreement between Bogoliubov and BCS pairings in describing the stable open shell nuclei was shown by taking the even Sn isotopes as the representatives. We have applied the RHFB theory with {the Gogny pairing force} to the study of Sn isotopes and $N=82$ isotones, and demonstrated that the RHFB theory with the finite-range Gogny force in the pairing channel can provide an appropriate quantitative description of both mean field and pairing correlation effects. In addition, better systematics from the stable region to the neutron-rich side are obtained with the inclusion of Fock terms, especially with the presence of $\rho$-tensor couplings which can eliminate artificial shell closures at 58 and 92. In fact, such improvements on systematics are meaningful for reliable explorations of exotic regions.

\begin{acknowledgments}

This work was supported by the Alexander von Humboldt Foundation, and Major State 973 Program 2007CB815000, as well as the National Natural Science Foundation of China under Grants No. 10435010, No. 10775004, and No. 10221003, {and by the DFG cluster of excellence \textquotedblleft Origin and Structure of the Universe\textquotedblright\ (www.universe-cluster.de)}.

\end{acknowledgments}


\end{document}